\documentclass[twocolumn,secnumarabic,superscriptaddress,amssymb,nofootinbib, aps, prb]{revtex4-2}
\usepackage{mathrsfs}  
\usepackage{graphicx}     
\usepackage{xurl}
\usepackage[plainpages=false,pdfpagelabels=true, linkcolor=cyan,breaklinks,colorlinks=true]{hyperref} 
\usepackage{xspace}
\usepackage{ulem}
\usepackage{xcolor}
\usepackage{amsmath}
\usepackage{enumitem}
\setitemize{leftmargin=1em}

\newcommand{\ybco}[1]{YBa\ensuremath{_2}Cu\ensuremath{_3}O\ensuremath{_{#1}}}
\newcommand{\Tc}{\ensuremath{T_{\rm c}}\xspace}

\begin{document}
	
	\title{Emergence of Fermi-liquid and BCS physics in overdoped cuprates}
	
	\author{B.~J.~Ramshaw}
	\email{bradramshaw@cornell.edu}
	\affiliation{Laboratory of Atomic and Solid State Physics, Cornell University, Ithaca, NY 14853, USA}
	\affiliation{Canadian Institute for Advanced Research, Toronto, Ontario, Canada}
	\author{Steven A. Kivelson}
	\email{kivelson@stanford.edu}
	\affiliation{Department of Physics, Stanford University, Stanford, California 94305, USA}

	\begin{abstract}
		Cuprates are the paradigmatic `unconventional' superconductors: their critical temperature is much higher than can be expected from phonon-mediated pairing; the superconducting gap has $d$-wave symmetry; and the normal-metallic state appears to be far from a conventional Fermi liquid. These and numerous other experimental facts have led to a consensus that the conventional theory --- the Fermi-liquid-based Bardeen--Cooper--Schrieffer (BCS) theory --- is the wrong starting point for understanding superconductivity in the cuprates. In this Perspective, we propose that, although underdoped cuprates do indeed require a different theoretical framework, there is a crossover with increasing doping to an overdoped regime in which a BCS-like approach is warranted (at energy scales of the order of the superconducting gap and below), provided that the various forms of disorder are accounted for. We summarize key experimental studies of the low-energy properties of overdoped cuprates, identify properties that are and are not compatible with this proposal --- and argue that features that are inconsistent with this approach can in fact be attributed to the expected effects of material disorder. Finally, we provide falsifiable predictions for the behaviour of an `ideal' (disorder-free) overdoped cuprate through which our approach can be tested. 
	\end{abstract} 
	
	\maketitle
	\section*{Introduction}
	
	For almost four decades, the search for a theoretical understanding of cuprate high-temperature superconductors has played a central role in condensed matter physics, in large part because these materials have been more intensively studied than any other correlated electronic material. Influential ideas that have germinated from such studies include the physics of quantum antiferromagnets (including possible spin-liquids) and doped antiferromagnets; novel forms of charge, spin, current and superconducting density wave orders; electronic analogues of various liquid crystalline phases; ways in which usual notions of metallic physics can break down (non-Fermi-liquid behaviours); bold new approaches to obtaining controlled (numerical) solutions of model problems --- such as the Hubbard model --- and a host of more phenomenological ideas focused on various experimental observations.
	
	A notable feature of of this body of research is the enormous effort made to achieve control over the materials involved. Results are now highly reproducible. However, this does not mean that the effects of disorder can be neglected. Most of the materials involved are {doped through chemical substitution}, so even in `perfect' materials, there is inherent disorder associated with the random configuration of dopant atoms. The measured properties of actual materials are certainly of interest, but it is also useful to ask what the features of a putative `ideal' --- that is, disorder-free --- cuprate would be. Central questions in the field concern the nature of the cuprate superconducting state itself: why is the critical temperature $T_{\textrm c}$ so high, and what limits $T_{\textrm c}$?; why do the phase diagrams of many families of cuprates exhibit a relatively similar superconducting dome-like feature?; and what can be learned from the cuprates about unconventional superconductivity in other material systems?
	
\begin{figure}
		\centering
		\includegraphics[width=\linewidth]{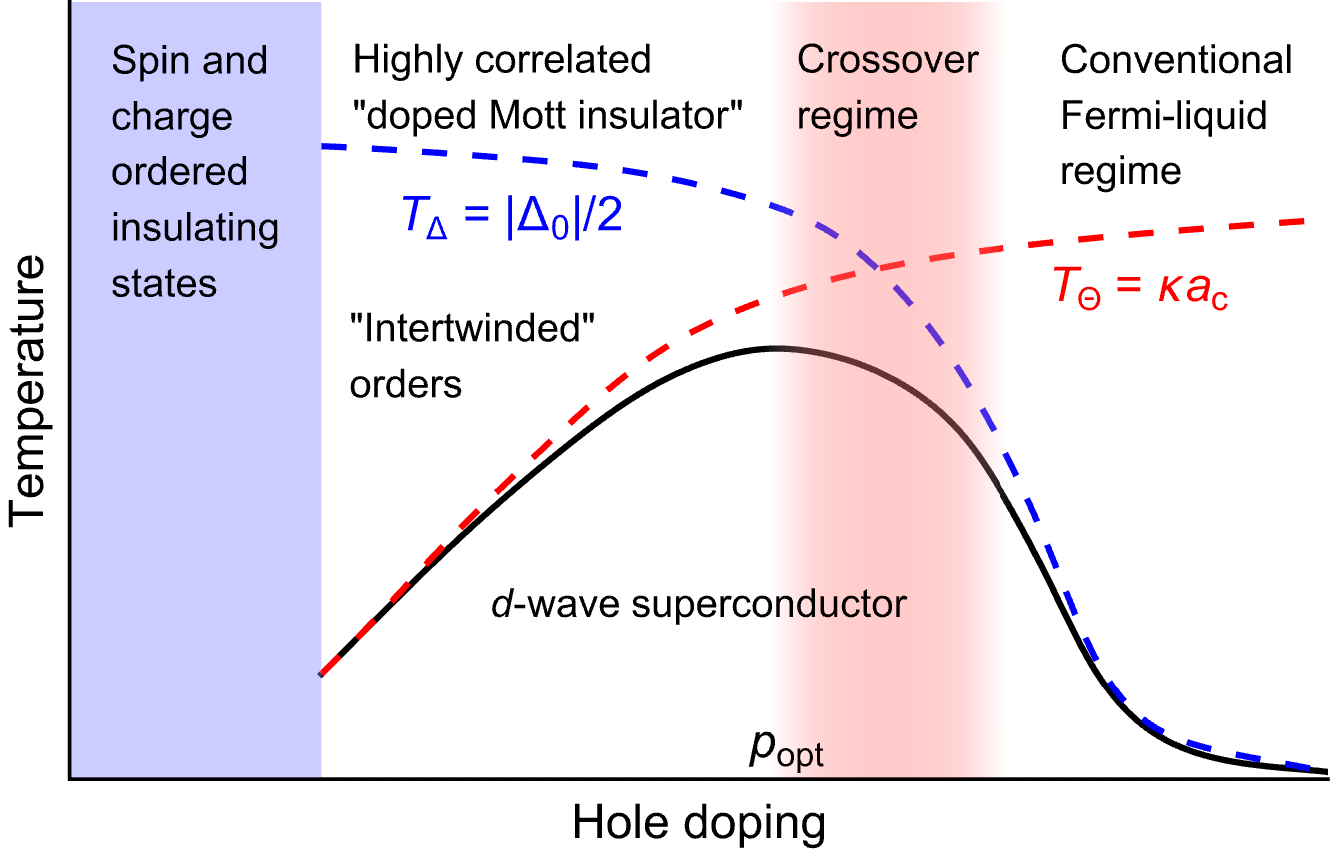}
		\caption{\textbf{ Conjectured phase diagram of an ideal disorder-free hole-doped cuprate 
				as a function of temperature $T$ and hole doping concentration $p$.} 
			The red region indicates a crossover 
			regime from an effective Fermi-liquid-like overdoped regime (the principle focus of this article) at larger $p$, to a highly correlated underdoped (`doped Mott insulator') regime  at smaller $p$.
			In contrast to the overdoped regime, the normal state at small $p$ is not well described in terms of weakly interacting quasiparticles, and various forms of intertwined order further complicate matters. 
			The black line indicates the superconducting transition temperature, $T_\textrm{c}$, 
			which traces a superconducting `dome' with a maximum at optimal doping, $p_\textrm{opt}$. It is bounded above by two energy scales that characterize the superconducting ground-state: $T_{\theta}\equiv \kappa a_{\textrm c}$ (red dashed line), the temperature scale at which thermal phase fluctuations alone would destroy superconducting order, where $\kappa$ is the zero-temperature superfluid stiffness and $a_\textrm{c}$ the mean distance between Cu--O planes (i.e. $\kappa a_\textrm{c}$ is the two-dimensional superfluid stiffness per plane); and the temperature scale $T_{\Delta}\equiv \Delta_0/2$ (blue dashed line), which is determined by the magnitude of the $T=0$ superconducting (pairing) gap,  $\Delta_0$. 
			At the lowest dopings (blue region), various phases exist with (at low $T$) long-range antiferromagnetic order --- either commensurate or incommensurate --- as do various forms of hole crystal, ranging from hole-Wigner crystals at the lowest doping to stripe-ordered insulators at higher doping. Essential considerations underpinning this phase diagram are detailed in Box 1. We note that aspects of this phase diagram are inconsistent with experiments, but we attribute those discrepancies to the effects of disorder.}
		\label{fig:PD1}
	\end{figure}

	In this Perspective, we consider the  evidence   that  the superconducting dome 
	in an `ideal' disorder-free cuprate arises from a
	crossover between a strongly correlated (and hence very complex) underdoped regime, where $T_{\textrm c}$ is intrinsically determined by phase ordering, to an overdoped regime, where it is possible to adopt a weak-coupling Bardeen--Cooper--Schrieffer (BCS) approach based on a Fermi-liquid normal state and a mean-field description of the superconducting state (but not, necessarily, invoking phonon-mediated pairing). 
	We present reasons to attribute to the effects of disorder those observed features of actual overdoped cuprates that are inconsistent with a weak-coupling description. Our viewpoint is summarized in \autoref{fig:PD1}, for which the essential assumptions are:
	
	
	\begin{itemize}
		\item{\bf $p < p_\textrm{opt}$} \\
		This `underdoped' regime may be best thought of as a lightly doped antiferromagnetic insulator.  Here, the pairing scale $|\Delta_0|$ is large compared to the superconducting transition temperature \Tc. Conversely, the energy scale that characterizes the stiffness to phase fluctuations \cite{emery_importance_1995}, $T_\theta$, is small compared to $|\Delta_0|/2$. Consequently, \Tc is the temperature at which phase fluctuations, rather than thermal pair-breaking, destroy long-range order. This hierarchy of scales, $|\Delta_0|/2 \gg \Tc \approx T_\theta$, is inverted from the usual weak-coupling case, and so probably reflects strong-coupling physics.  Indeed, in this regime of doping there is direct evidence for various types of `intertwined' density-wave orders (which are generically strong-coupling effects) and a host of fluctuation phenomena that have been linked under the name `pseudo-gap phenomena'. There is no generally accepted theory that describes the rich physics of the underdoped regime.
		
		\item{\bf $p > p_\textrm{opt}$} \\  
		On the overdoped side of the superconducting dome, the familiar hierarchy present in BCS superconductors --- 
		$T_\theta \gg \Tc \approx |\Delta_0|/2$ ---
		is respected.
		Confirming this behaviour in the limit of vanishing disorder would constitute strong evidence of the applicability of a weak-coupling theory. Correspondingly, the normal state, with superconductivity suppressed either by increasing temperature or magnetic field, should be well captured by the usual Fermi-liquid theory of metals.  
		
		\item{\bf $p\sim p_\textrm{opt}$} \\ 
		For a range of dopings in the vicinity of optimal doping, $p=p_\textrm{opt}$, there is a  regime where both $|\Delta_0|/2$ and $T_{\theta}$ are comparable to \Tc. This is a crossover regime, where presumably no single perspective can easily capture all of the physics.
		
	\end{itemize}
	
	Our primary purpose is to critically assess the validity of a weak-coupling approach to the overdoped cuprates. (Although most theoretical studies of the cuprates have emphasized more exotic, strong-coupling aspects of the problem, there is a subset of studies that overlap either in some details or in perspective with the present approach \cite{myreview,chakravarty_hidden_2001,laughlin_hartree-fock_2014,Peter1,Peter2,scalapino0,scalapino1,scalapino2,dunghaiandseamus,normanperspective}.)
	Specifically, we identify the experimental evidence that supports our viewpoint and, perhaps more importantly, the observations that clearly contradict it. For example, several families of cuprates exhibit the `boomerang effect', where the superfluid phase stiffness (written in temperature units as $T_\theta$) decreases as the doped hole concentration ($p$) increases, such that $T_\theta$ remains comparable to $T_{\textrm c}$ throughout the overdoped regime and thus `turns around' when plotted parametrically as a function of $p$ on a $T_\theta$--\Tc plot \cite{uemura_universal_1989,UemuraBoomerang,bozovic_dependence_2016,lemberger}. Moreover, when superconductivity is suppressed with a high magnetic field, strange metal behaviour (as opposed to the expected Fermi-liquid behaviour) is seen throughout the overdoped regime \cite{cooper_anomalous_2009,proust_heat_2002}. 
	We will discuss reasons to think that these features are the consequence of intrinsic, substitutional disorder. 
	
	The conceptual underpinnings of the our approach are distinct from those underlying two-fluid phenomenology \cite{ayres_incoherent_2021}: whereas we identify the various  properties of  overdoped materials that are inconsistent with a Fermi-liquid or BCS description as `extrinsic' (that is, vanishing in the zero-disorder limit), two-fluid phenomenology interprets them as consequences of unspecified strong correlation effects that lead intrinsically to the coexistence of coherent and incoherent fluids of electronic excitations. Our approach is also fundamentally different from the widely discussed perspective in which optimal doping is associated with a  quantum critical point whose influence dominates large portions of the phase diagram\cite{wangchubukov,tailleferreview,subirreviewqcp}. Although we do not preclude the existence of one or more quantum critical points below optimal doping, one outcome of our discussion will be that, under the assumption that the weak-coupling analysis is reasonable for hole doping down to values comparable to optimal doping, a putative quantum critical point 
	near optimal doping does not play a major role in the pairing mechanism.
	
	There is a lot riding on this. Admittedly, without addressing the crossover to the strongly correlated physics on the underdoped side of the phase diagram (\autoref{fig:PD1}), we cannot claim to fully understand superconductivity at optimal doping, nor why $T_{\textrm c}$ drops with underdoping. However, if our approach is verified, it would provide an understanding of  high-temperature superconductivity from the overdoped side. As the cuprates are the best characterized of all unconventional superconductors, even this partial understanding would have broad implications for the field.

	\section*{The BCS superconductor}
	\label{se:bcs}
	
The term `BCS superconductor' immediately invokes the theory of phonon-mediated pairing that is the quintessentially successful theory of conventional $s$-wave superconductors. The cuprates, however, are `unconventional' $d$-wave superconductors.  So it is important to stress that here we use `BCS superconductor' in a broader sense, encompassing all cases in which the superconducting state emerges as a weak-coupling instability of a Fermi-liquid normal state, independent of whether or not phonons play a role in the mechanism.

\subsection*{Conventional superconductors and BCS theory}

The properties of crystalline (`clean') conventional superconductors can be derived from the electron--phonon problem using the Migdal--Eliashberg approximation. Unlike the original BCS theory, which abstracts away the pairing mechanism with a generic attractive potential, this approximation explicitly keeps both the electrons and the `bosonic glue' of the phonons.

At temperatures $T>\hbar\omega_D$ (where $\omega_D$ a characteristic phonon frequency, such as the Debye frequency, and $\hbar$ is the reduced Planck constant), the quasiparticle lifetime in a metal is $\hbar/\tau = 2\pi \lambda_0 T$, where $\lambda_0$ is the dimensionless (unrenormalized) electron--phonon coupling constant. It is not immediately obvious to what extent Fermi-liquid theory applies at these temperatures, where quasiparticles (in the Landau sense) are at best marginally well defined \cite{prange_transport_1964}. However, if one is only interested in treating the problem at temperatures and energies that are small compared to  $\hbar \omega_D$, then it is possible to introduce (as was done by Bardeen, Cooper and Schrieffer) a simpler low-energy description of the system in terms of well defined Landau quasiparticles with a weak residual interaction proportional to the dimensionless parameter $\lambda-\mu^\star$. Here, $\lambda \sim \lambda_0/(1+\lambda_0)$, and $\mu^*$ are renormalized versions of the dimensionless electron--phonon and electron--electron interactions. Formally, this `effective field theory' can be derived\cite{1992hep.th...10046P,shankar,carlsonreview} from the original problem by integrating out degrees of freedom with energies higher than $\sim \hbar \omega_D$, so that $\hbar\omega_D$ appears only as the high-energy cut-off of the theory. (There are a couple of subtleties concerning the BCS prescription for integrating out the phonons: as there are always acoustic phonons, it is not literally the case that all the phonons are integrated out, but these modes do not couple strongly to the electrons and can thus be neglected; and, even at low $T$, there can be dynamical effects --- such as relaxation of non-equilibrium populations of quasiparticles --- that cannot be well understood in terms of the effective field theory and require a theory in which phonons are included explicitly.)

An essential piece of evidence that the mechanism of pairing is a phonon-induced attraction between electrons comes from analyzing the energy dependence of the gap function, $\Delta(\vec k,\omega)$, at energy scales $\hbar \omega\sim\hbar \omega_D$ --- that is, at energies where Landau quasiparticles are not sharply defined. In the context of our upcoming discussion of the overdoped cuprates, a take-away message is that the lack of a conventional quasiparticle description at energies much greater than \Tc does not preclude a weak-coupling BCS approach at 
{energies $\sim \Tc$ and below.}

For additional insight, consider a conventional superconductor from a low-$T$, low-energy, empirical perspective and imagine trying to infer something about the pairing mechanism. To begin with, well-defined quantum oscillations arise when superconductivity is suppressed by a magnetic field, and thus the normal state is a Fermi-liquid metal. Three facts then indicate that a mean-field description is warranted: first, the gap vanishes at $\Tc$; second, $\Tc$ is determined by the quasi-particle gap $\Delta_0$, that is, \Tc is equal to $|\Delta_0|/2$ (up to corrections that are never more than 20-30\% effects); and, third, thermodynamic quantities such as the specific heat show mean-field-like behaviour. Finally, the measured gap function, $\Delta(\vec k,\omega)$, is only weakly dependent on the wavevector $\vec k$, suggesting that pairing is induced by an effective attractive interaction that is local (short-range) in space. All of these phenomena are consistent with the underlying assumptions of BCS theory. 

More detailed information about the pairing mechanism is somewhat hidden in this BCS approach. In principle, the attractive interaction could come from something other than phonons. Indeed, this is
the sense in which we are suggesting that overdoped cuprates can be described using BCS theory:
it is clear that the pairing interaction in the cuprates
is not, primarily, due to phonons, and yet we will argue that a BCS perspective is warranted. (It is also not entirely true that there are no signatures of the mechanism; in conventional superconductors, the importance of phonons can (often) be derived from the isotope dependence of the cut-off in the theory, an observation that played an essential role in the acceptance of BCS theory.) 

\subsection*{Negligible fluctuations}

BCS theory is a mean-field theory and is therefore only valid when fluctuation effects are negligible. Close enough to \Tc, universal critical fluctuations must arise, meaning that BCS theory is never applicable {over some temperature range}. However, in conventional superconductors, significant fluctuation effects are confined to extremely narrow regions around \Tc (of order $10^{-8}$\Tc \cite{blatter_vortices_1994}). 

Although there are many possible fluctuation effects that can give corrections to BCS theory, particularly important in the cuprate context are phase fluctuations.
In the superconducting state below \Tc, the low-energy physics is universally captured by the non-linear sigma model, with free energy, $F$, given by
\begin{equation}
	F = \frac {\kappa(T)}2\ \int d\vec r  \ \left|\vec \nabla \theta -\frac {2e}{\hbar c} \vec A\right|^2
\end{equation}
where $\kappa$ is known as the superfluid stiffness, $\theta(\vec r)$ is the local phase of the superconducting order parameter, and $\vec A$ is the electromagnetic 
vector potential; $e$ is the electron charge and $c$ is the speed of light in vacuum.
Importantly, $\kappa$ is a measurable quantity:  in three dimensions, $ \kappa(T) = [(\hbar c)^2/16\pi e^2]\ \lambda^{-2}(T)$, where $\lambda$ is the London penetration depth. We define an energy scale associated with phase fluctuations as 
\begin{equation}
	k_{\rm B} T_{\theta}(T)\equiv \kappa(T) L^{d-2}
	\label{Ttheta}
\end{equation}
where $L$ is an appropriate microscopic length scale that provides an ultraviolet cutoff for the theory, $d$ is the number of dimensions and $k_{\rm B}$ is the Boltzmann constant. We will adopt the notation $T_\theta \equiv T_{\theta}(T=0)$. Of particular interest here is that when  $d=2$, $\kappa$ has units of energy. For a layered system of weakly Josephson-coupled superconducting planes, the appropriate length scale $L$ is the spacing between planes, {$a_c$.  (More generally, $L$ is the larger of 
	$a_c$} and the superconducting coherence length, $\xi_c$, in the interplane direction.
One consequence of this is that, close enough to \Tc, the physics must universally cross over to three-dimensional (3D) criticality, in which case $L \sim \xi_c$.)

The meaning of this energy scale is that thermal phase fluctuations are dilute (and hence in certain senses negligible) as long as the temperature is much less than  $T_\theta(T)$,  and that they have an order-one effect only when the temperature is of order $T_{\theta}(T)$ or greater. 
Close enough to \Tc, phase fluctuations are always 
important.  In two dimensions, as $T \to \Tc$ from below, $T_\theta(T)
\to (2/\pi)$ \Tc 
(that is, there is a universal drop of the phase stiffness at $\Tc$). In three dimensions (to the extent that  gauge-field fluctuations can be ignored), $T_\theta$ goes to zero 
as $T\to \Tc$ {from below}, with universal 3D XY critical exponents. 
However, in both cases, if at low temperatures $T_\theta \gg \Tc$, phase fluctuations are expected to be important only in a range of temperatures of order $\Delta T \sim \Tc (\Tc/T_\theta)$. Thus, even in the case of a 
film of a conventional, clean superconductor of microscopic thickness $W$, 
universal two-dimensional XY behaviour is essentially unobservable because $T_\theta$ is of the order of the Fermi temperature ($T_\theta \sim E_F (k_F W)\gg \Tc$). 

\section*{A BCS approach to the overdoped cuprates}

Because the cuprates are strongly correlated, the theoretical route from the microscopic physics of copper and oxygen orbitals to an appropriate low-energy effective model is difficult and will not be addressed here. Instead, we assume that there exists an intermediate energy scale that is large compared with $T_{\textrm c}$ --- possibly of the order of the exchange energy in the parent antiferromagnetic insulator, $J \approx 1500$ K \cite{j_birgeneau_magnetic_2006}. Below this scale, the electronic properties can be understood on the basis of an effective model of well-defined emergent quasiparticles with relatively weak residual interactions.  The quasiparticle properties may be strongly renormalized, and the residual interactions can encode the effects of important high-energy physics, especially the existence of local moments on the copper sites. Importantly, however, because the effective theory is obtained by integrating out high-energy degrees of freedom, the residual interactions can be taken to be short-ranged, and both the interactions and the quasiparticle renormalizations must be smooth functions of microscopic parameters, such as the doping concentration, $p$. The existence of quantum oscillations in some overdoped cuprates (which we will discuss in more detail later), constitutes strong evidence that this assumption is reasonable, at least at low temperatures.

The BCS self-consistency condition is derivable as an implicit expression for the gap function, $\Delta(\vec k)$ (which has no explicit $\omega$ dependence because we are dealing with an effective Hamiltonian), as a function of the effective interaction, $\tilde V(\vec r)$. This condition is usually expressed as a convolution over momentum, $\vec q$, of $V(\vec q)$ --- the Fourier transform of $\tilde V(\vec r)$ --- and an integrand that depends on the band dispersion, $\epsilon(\vec k)$, and the quasiparticle energy,
\begin{equation}
	E(\vec k) \equiv \sqrt{|\epsilon(\vec k) - \mu|^2 +|\Delta(\vec k)|^2},
\end{equation}
where $\mu$ is the chemical potential.  
{It} is also useful to look at this relation in real-space:
\begin{equation}
	\tilde \Delta(\vec r) = -\tilde V(\vec r)\ \tilde \phi(\vec r),
	\label{BCS}
\end{equation}
where $\tilde \phi(\vec r)$ 
is the  pair wave-function,
\begin{equation}
	\tilde \phi(\vec r) = \int \frac {d\vec k}{(2\pi)^d} \ e^{i\vec k\cdot \vec r}\ \frac {\Delta(\vec k)}{2E(\vec k)} \tanh\left[\frac{E(\vec k)} {2T}\right]  \ F\left[\epsilon(\vec k)\right],
\end{equation}
where $F(\epsilon)$ implements a high-energy cutoff of the theory, if needed.

One feature of BCS theory that is often overlooked is that the form of the gap function is almost entirely determined by the form of the effective interaction. This follows from equation \ref{BCS} and the fact that $\tilde \phi$ is much longer  ranged (more slowly varying as a function of $|\vec r|$) than $\tilde V$. This explains why the gap function $\Delta(\vec k)$ is typically approximately momentum independent in conventional superconductors {(higher harmonics of the gap function require longer-range interactions)}. It also accounts for the fact that, although the magnitude of $\Delta(\vec k)$ changes with temperature and vanishes at \Tc, its $\vec k$ dependence remains roughly constant --- even in unconventional superconductors.  For the case of a $d$-wave superconductor, relevant to the cuprates, 
\begin{equation}
	\Delta(\vec k) = \Delta_0\ d(\vec k) \ f(\vec k)
	\label{equation6}
\end{equation}
where $\Delta_0$ characterizes  the magnitude of the gap function; 
$d(\vec k)\equiv 
(1/2)[\cos(k_x) - \cos(k_y)]$ is mandated by the $d$-wave symmetry; and $f(\vec k)$ (which determines the range of $\tilde \Delta$)  transforms trivially under all point group transformations and is normalized such that  $f(0,\pi)=f(\pi,0)=1$.  

In the case of the shortest-range interactions allowed by symmetry, $f(\vec k)=1$. This corresponds to a gap function that is non-zero only on nearest-neighbour bonds. To be explicit, we will treat the case of a density--density interaction, $\tilde U(r)$, and a spin--spin interaction (exchange coupling), $\tilde J(r)$, which vanish beyond nearest-neighbour distance: $\tilde U(r)=\tilde J(r)=0$ for $r>1$ {(where $r$ is measured in units of the lattice parameter)}. In this case, whenever a non-zero $\Delta_0$ exists, the BCS self-consistency relation can be written as
\begin{equation}
	1 = -\tilde V\int \frac {d\vec k}{(2\pi)^d} \ \frac {|d(\vec k)|^2}{2E(\vec k)} \tanh\left[\frac{E(\vec k)} {2T}\right]  \ F\left[\epsilon(\vec k)\right]
	\label{linearizedgap}
\end{equation}
with $\tilde V= -\left[3 \tilde J(1)-4\tilde U(1)\right]$. (Because we have assumed $d$-wave symmetry, the onsite interaction $\tilde U(0)$ does not play any role in pairing.) This equation applies only for the case $\tilde V < 0$, and below \Tc. Note that, although there is a broad consensus that the dominant contribution to $\tilde V$ is associated with incipient (short-range) antiferromagnetic interactions ($\tilde J(1)>0$), from the present perspective this cannot be distinguished from the effects of a short-range  density--density attraction ($\tilde U(1) <0$) that could arise from interactions with appropriate high-energy optical phonon modes. Equation \ref{linearizedgap} is typically viewed from a bottom-up perspective as a route to compute $\Delta_0(T)$ from a given (assumed known) effective interaction;  below, we will invert this logic, and use measured values of $\Delta$ to infer the nature of the effective interaction, $\tilde V$.

Within BCS theory, \Tc vanishes only when $\tilde V$ becomes positive. Because the dependence of the effective parameters on microscopic parameters is analytic by assumption, in the present context this means that if there is (in the absence of disorder) a critical value of the doping, $p_{max}$, beyond which \Tc vanishes, then 
\begin{equation}
	\tilde V \approx -\tilde V_0\ (p_{max}-p), 
	\label{tildeV}
\end{equation}
where $\tilde V_0>0$. Consequently, \Tc $\sim e^{-\alpha (p_{max}-p)^{-1}}$ where $\alpha \propto |1/\tilde V_0|$ is a dimensionless constant. In other words, if the pairing interaction changes sign at a critical doping, \Tc will decrease exponentially on approach to that doping, rather than cutting off sharply as it does at the edge of a `dome'. This is the basis of the overdoped `tail' shown in \autoref{fig:PD1}.

\subsection*{Effects of disorder on a BCS $d$-wave superconductor} 

It might naively be thought that superconductivity in the weak-coupling (BCS) limit requires extremely clean systems, such that the quasiparticle mean-free-path, $\ell$, is large compared to the clean-limit coherence length, $\xi_0$.  It is a special property of conventional $s$-wave superconductivity, known as Anderson's theorem, that both \Tc and $\Delta_0$ are largely independent of $\xi_0/\ell$. Generically, disorder only notably affects \Tc  when it is strong, that is, when $k_F \ell \sim 1$.
However, 
it follows straightforwardly from the effective field-theoretic approach of Abrikosov and Gorkov that for $d$-wave and other superconductors with a sign-changing gap, the pair-breaking effect of disorder results in the elimination of the superconducting state when $\xi_0/\ell$ exceeds a critical value of order 1.  

Indeed, there are some features of the resulting theory that resemble observed properties of overdoped cuprates \cite{peteruniform1,peteruniform2,peteruniform3}. For one thing, Abrikosov--Gorkov theory gives an explicit expression for how \Tc is suppressed with increasing $\xi_0/\ell$, in which \Tc vanishes linearly as $\xi_0/\ell$ approaches  the critical value. Moreover, the zero-temperature superfluid stiffness, $\kappa$, vanishes in proportion to \Tc
as $\Tc\to 0$. However, this does not indicate that phase fluctuations become important (and BCS theory therefore inaccurate).  Specifically, in
the 3D (or quasi-two-dimensional) version of the expression in equation \ref{Ttheta} --- in which the length scale, $L$, is the $c$-axis coherence length, $\xi_c$ --- because $\xi_c$ diverges as $\Tc\to 0$, it follows that $T_\theta/\Tc$ also diverges. Even in the fully two-dimensional case, in which $L=a_c$, it follows that 
\begin{equation}
	T_\theta/\Tc \sim (k_F\ell)  \sim (k_F\xi_0) \gg 1 \ {\rm as} \ \Tc\to 0.
\end{equation}
As far as we know, accessing a regime in which $T_\theta/\Tc \lesssim 1$ requires something beyond BCS theory.

Abrikosov--Gorkov theory treats disorder in 
an effective medium approximation. This is expected to be extremely accurate in the usual situation in which $\xi_0 \gg L_{\textrm {dis}}$, where $L_{\textrm {dis}}$ is the correlation length of the disorder --- a hierachy that applies to many unconventional superconductors including the iconic example of Sr$_2$RuO$_4$ (ref.\cite{mackenziedisorder}). 
The cuprates are special, even  in the context of BCS mean-field theory, in that they are $d$-wave and that they have a relatively small coherence length, with $\xi_0 \sim $ 2 or 3 times the lattice constant.  In particular, $\xi_0$ is comparable to $L_{\textrm {dis}}$, which is probably  
of order $p^{-1/2}$ times the lattice constant, and \cite{tranquadaNonmetalStrangeMetal2024};
consequently
mesoscopic fluctuations in the local value of $\ell$  inevitably lead to large variations in the local value of \Tc. Therefore, near the edge of the superconducting dome the effects of disorder are expected to be amplified, leading to an effectively granular state with locally superconducting puddles emersed in a metallic background.  Various numerical and analytic studies have confirmed this expectation \cite{dunghaiandme,Peter2,Peter1,sulangi}. Importantly, 
the inevitable consequence is a more rapid suppression of $T_\theta$ than of the spatially averaged \Tc. Thus, near the edge of the superconducting dome, phase fluctuations inevitably become important and lead to a breakdown of the mean-field description.

There are additional considerations that we expand upon in the Supplementary Information. There, we review the difference between a crossover and a phase transition in the context of the assumed crossover in \autoref{fig:PD1}, which is essential to the present analysis in that it justifies treating the mechanism of superconductivity in the overdoped regime without needing to account for the behaviour in the underdoped regime, despite the fact that they are smoothly (adiabatically) connected to one another (Supplementary Information section I). We discuss the relation between our approach and those that treat the existence of a putative quantum critical point at $p=p^\star$ under the superconducting dome as the central organizing feature of the cuprate phase diagram (Supplementary Information section II). Finally, we expand upon the tangible differences between homogeneous `dirty $d$-wave' and inhomogeneous, self-organized granularity as the  cause of the suppressed superfluid density on the overdoped side of the phase diagram (Supplementary Information section III).

\section*{Experiments on overdoped cuprates}

There is a body of experimental evidence that contains useful information on the applicability of the BCS perspective in overdoped cuprates. Here, we first summarize the experimental observations that most clearly support our proposed approach, and then those that are inconsistent (or apparently inconsistent) with it. We will argue that features inconsistent with the BCS perspective can be plausibly understood as  fluctuation effects arising from the inhomogeneous electronic structure of a disordered $d$-wave superconductor with a short coherence length --- that is, in the absence of disorder, the features that are inconsistent with BCS theory would be absent.

We focus on experiments relevant to overdoped cuprates, and touch on underdoped experiments only when the comparison is helpful.
	To further focus the discussion, we consider only hole-doped cuprates. (As mentioned below, the antiferromagnetic correlations in the electron-doped cuprates that persist to optimal doping, and beyond, might complicate the application of the present analysis to these materials \cite{greenereview}). We present 
	what we consider to be the strongest evidence --- for example, we use quantum oscillations to argue for a Fermi-liquid ground state rather than  power laws in various transport coefficients --- and we restrict our attention to `low energy' properties, that is to say, those that reflect excitations with energies less than some intermediate scale,  
	which we take somewhat arbitrarily to be $\sim 0.1$ eV 
	or, in other words, a scale of the order of the exchange energy $J$ in the undoped parent materials.
	We also focus on a subset of materials: \ybco{6+\delta} (YBCO), the electronically cleanest cuprate; La$_{2-x}$Sr$_x$CuO4 (LSCO), the cuprate with the most comprehensive data sets, including in the low-temperature normal state accessed by magnetic fields; Tl$_2$Ba$_2$CuO$_{6+\delta}$ (Tl2201), the cleanest cuprate that can access  the overdoped regime; and Bi$_2$Ba$_2$CaCu$_2$O$_{8+\delta}$ (Bi2212), which has the best angle-resolved photoemission (ARPES) and scanning tunnelling microscopy (STM) data. 
	
	One issue that arises when comparing results reported in different studies on multiple families of cuprates is that the doped hole concentration, $p$, is  typically inferred  rather than measured.  Indeed, often it is estimated on the basis of an assumed universal form of the $\Tc(p)$ dome. As we will conclude that the termination of the  dome is highly sensitive to the character and nature of the disorder, we have not adopted this approach.  Instead,  we take (wherever possible) the value obtained from the  area enclosed by the Fermi surface, measured using low-energy ARPES or from quantum oscillations.

\subsection*{The Fermi liquid ground state}

For overdoped cuprates, quantum oscillations, angle-dependent magnetoresistance and ARPES imply a Fermi surface similar to what is found in local-density approximation (LDA) calculations.

\subsubsection*{Experiments in support of a Fermi liquid ground state.}

Quantum oscillations --- arising from Landau quantization of quasparticle orbits in a magnetic field --- are the gold standard for demonstrating a Fermi-liquid ground state. At high enough magnetic fields to quench superconductivity, quantum oscillations 
are observed in Tl2201 at dopings higher than $p=0.27$, corresponding to $\Tc \leq 26$~K \cite{rourke_detailed_2010,vignolle_quantum_2008}. Quantum oscillations are not observed for $p<0.27$, possibly owing to the presence of two-dimensional, short-range, correlated charge-density-wave order \cite{tam_charge_2022,gannot_fermi_2019}.

The quantum oscillations demonstrate the presence of a Fermi surface; moreover, it is a single, hole-like, cylindrical Fermi surface. The measured areas correspond to hole densities of $p = 0.27$ for $\Tc = 26$ K and $p = 0.3$ for $\Tc = 10$ K \cite{rourke_detailed_2010}, and are consistent with those inferred from angle-dependent magnetoresistance \cite{hussey_coherent_2003}, and  ARPES \cite{plate_fermi_2005}. The temperature dependence of the quantum oscillations fits the Lifshitz--Kosevich form of a Fermi liquid very well from $0.3$ K to $3$ K, yielding a cyclotron effective mass of approximately $m^{\star} = 5$ $m_e$, where $m_e$ is the bare electron mass.

The quasiparticle scattering rate is $\hbar/\tau=2.4$ meV. Note that this is the not the transport scattering rate --- quantum oscillations are sensitive to forward-scattering processes that do not affect the resistivity. The angle dependence of the quantum oscillations is consistent with a $g$-factor of $2$ (ref. \cite{rourke_detailed_2010}), similar to underdoped YBCO (ref. \cite{ramshaw_angle_2011}).

Although the Fermi-surface geometry agrees very well with LDA (ref. \cite{singh_electronic_1992}), a bandwidth renormalization of a factor of three is required to match the cyclotron mass with that calculated from LDA. This is similar to the mass renormalizations found in Sr$_2$RuO$_4$ (ref. \cite{bergemann_quasi-two-dimensional_2003}). The same bandwidth renormalization is found over hundreds of meV by ARPES \cite{peets_tl2ba2cuo6_2007}, which demonstrates that the quasiparticle mass renormalization arises predominantly through eV-scale electron--electron interactions, and not from interactions with a low-energy boson \cite{rourke_detailed_2010}. 
Moreover, interpreting the temperature dependence of the amplitude of quantum oscillations 
in terms of only an effective mass assumes that corrections from an energy-dependent self-energy are small \cite{shoenberg_magnetic_1984}. The absence of any deviation from the standard Lifshitz--Kosevich form in Tl2201 implies that there are no strongly energy-dependent renormalizations  (as there would be in the marginal Fermi liquid \cite{shekhter_thermodynamic_2017}), even with a \Tc as high as 26 K.

\subsubsection*{Experiments that challenge a Fermi-liquid ground state.}

The strongest arguments against a Fermi-liquid ground state stem from observations of various transport anomalies at low $T$ when superconductivity is suppressed by a high magnetic field.  Of these, the most dramatic is the $T$-linear resistivity extending to low temperatures in a range of doping from $p_{opt}$ to the overdoped edge of the superconducting dome, $p_{max}$ (ref. \cite{cooper_anomalous_2009}).  (Note that here we are not discussing the extension of $T$-linear resistivity to very high temperatures that is observed near optimal doping, which
has long been argued to indicate the existence of
a non-Fermi-liquid ground state at or near a special doping, $p^*$).

Specifically, $T$-linear resistivity has been reported \cite{cooper_anomalous_2009} out to the very edge of the superconducting dome in LSCO --- when superconductivity disappears, so does the $T$-linear resistivity. These measurements are consistent with earlier observations in Tl2201 (ref. \cite{proust_heat_2002}).  
It has also been suggested that the approximate correspondence between the ranges of $p$ for these observations implies that the scattering responsible for $T$-linear resistivity is also responsible for superconducting pairing. 
From a somewhat different perspective, 
the observation of $B$-linear magnetoresistance near optimal doping in LSCO (ref. \cite{giraldo-gallo_scale-invariant_2018}), and later at higher dopings in Tl2201 and Bi2201 (ref. \cite{ayres_incoherent_2021}), has been interpreted as evidence for scale invariance, as might be expected near a quantum critical point.  

\subsubsection*{Reconciling results on the Fermi-liquid ground state.}

There have been several theoretical proposals attempting to reconcile the unambiguous evidence from quantum oscillations for the existence of a Fermi liquid with the observed transport anomalies. As discussed above, even a high-quality `uniformly doped' sample will appear heterogeneous on short enough length scales, owing to the statistical distribution of dopant locations. This can lead to an electronically inhomogeneous state that consists of dilute superconducting `puddles'  that might survive even to the high fields at which superconducting long-range coherence is destroyed; it has been demonstrated that scattering from the  quantum phase fluctuations on such puddles can give rise to a $T$-linear resistivity \cite{bashan_extended_2025}. 

Experimental evidence for such a scenario comes from overdoped Tl2201, where increasing the magnetic field from 10 to 18 tesla reduces the $T$-linear slope by at least a factor of two (ref. \cite{mackenzie_normal-state_1996}). 
Alternatively, the possibility that  anomalous scattering might be associated with local, nearly-quantum-critical patches of some other order --- possibly associated with a  disorder-broadened antiferromagnetic quantum critical point --- has been proposed in a series of related papers\cite{subir1,avishkar1,esterlisschmalian,subirandmax,berg1}. 

For the present purposes, what is essential is that these proposals imply that an extrinsic (disorder-related) form of anomalous scattering is responsible for the non-Fermi-liquid transport.  Our specific prediction is that the  $T$-linear term will be absent in 
less-disordered overdoped 
cuprates, such as stoichiometric YBCO doped 
through application of hydrostatic pressure \cite{alireza_accessing_2017}.

\subsection*{Negligibility of superconducting fluctuations}
 When the phase ordering scale, $T_\theta$,  is comparable to \Tc,  phase fluctuations necessarily are an order-one effect; 
this implies a breakdown of BCS mean-field theory.  This situation applies in all underdoped cuprates.  The issue is more complex in the overdoped cuprates.

\subsubsection*{Experimental evidence that fluctuations are insignificant}

Although in LSCO and Tl1201 
the superfluid density (or more precisely  $T_\theta$)
decreases with increasing $p$ past optimal doping (the `boomerang effect'), in \ybco{x} it increases monotonically as a function of $p$ across the entire accessible range of doping
\cite{pereg-barnea_absolute_2004}. Particularly notable is the fact that
$T_\theta$ increases \cite{baglo_detailed_2014} by $50\%$ from its value at $p = 0.16$, where $\Tc = 94$ K, to $p = 0.19$, where $\Tc = 88$ K (\autoref{fig:phase_stiff_all}). This departs markedly from the Uemura relationship, where phase stiffness is proportional to the superconducting \Tc \cite{uemura_universal_1989}.  

\ybco{7-\delta} does, in fact, exhibit an unambiguous temperature range over which phase fluctuations are 
{significant} \cite{breit_evidence_1995, pasler_3d-xy_1998,noauthor_bounding_nodate,kamal_penetration_1994}. The measured fluctuations are consistent with 3D-XY critical fluctuations \cite{pasler_3d-xy_1998,kamal_penetration_1994}, and they extend to temperatures of order 10\% above \Tc at optimal doping ($\delta\approx 0.07$) but over a smaller range of $T$ for overdoped \ybco{7} \cite{breit_evidence_1995}. We conclude that, although 
phase fluctuations in overdoped \ybco{7} are much larger than in conventional superconductors, where the fluctuation regime is $\approx 10^{-9}\Tc$ (ref. \cite{blatter_vortices_1994}),  
they play a relatively minor role in determining \Tc.

\subsubsection*{Experimental evidence of strong phase fluctuations.}

Mutual-inductance measurements of the penetration depth on thin-film LSCO (ref. \cite{bozovic_dependence_2016}), as well as measurements of the same quantity using muon spin resonance on single-crystal Tl2201 (ref. \cite{uemura_magnetic-field_1993}), show a vanishing superfluid phase stiffness approaching the edge of the superconducting dome. Indeed, the ratio of $\Tc/T_\theta$ behaves quite similarly on the underdoped and overdoped sides of the superconducting dome \cite{bozovic2}. Large phase fluctuations  have also been inferred from the width and height of the jump in specific heat at \Tc in Tl2201 (ref. \cite{wade_electronic_1994}), and from the onset of an apparent diamagnetic signal in measurements of the Nernst effect in overdoped LSCO, Bi$_2$Sr$_2$CuO$_{6+_\delta}$ (Bi2201) and Bi2212 (refs \cite{wang_onset_2001,wang_nernst_2006}). 
The implication of these experiments in terms of phase fluctuations contradicts a BCS scenario.

\subsubsection*{Evidence that large phase fluctuations are extrinsic.}

There are three classes of mechanism that could suppress the macroscopic phase stiffness of overdoped cuprates relative to the bare Drude weight: strong-coupling physics, pair-breaking due to scattering \cite{peteruniform1,peteruniform2,peteruniform3}, and mesoscopic inhomogeneity (on scales of the order of the coherence length). The first mechanism is intrinsic, and the latter two are extrinsic in the sense that they disappear in the limit of  
vanishing disorder. While all three effects may be present to some degree, but 
we argue that extrinsic effects are dominant. Further, we suggest that mesoscopic inhomogeneity provides a more compelling explanation for several experimental phenomena, although it should be noted that the debate over the relative importance of pair breaking versus inhomogenaity is ongoing \cite{peteruniform1,peteruniform2,peteruniform3} (see Supplementary Information for further discussion).

Needless to say, macroscopic inhomogeneities in the dopant concentration or crystal structure will inevitably produce electronic inhomogeneities on a similar scale. This sort of inhomogeneity is thought to be a feature of strongly overdoped LSCO, where it leads to an inhomogeneous (`Swiss cheese') superfluid response \cite{swiss1,swiss2}. (There is also evidence that the superfluid response of Tl2201 is inhomogeneous on the scale of a few hundred angstroms \cite{ioffemillis}.) But even in  macroscopically homogeneous materials, we expect that alloy disorder will lead to spatial variations of the local electronic structure on more microscopic scales.  Direct evidence of large-amplitude local electronic inhomogeneity comes from an analysis of the inhomogeneous broadening of the spectra from nuclear quadruople resonance (NQR) of various cuprates \cite{haas,singerNQREvidence2002,singerNMRStudy2005}; almost all cuprates exhibit variance in the local  hole concentration measured on Cu sites that is comparable to the total width of the superconducting dome. 
There are three exceptions, in which the NQR linewidth is 
much smaller than in other cuprates: highly-ordered YBa$_2$Cu$_3$O$_{6.50}$ Orth-II (ref. \cite{wu_incipient_2015}) and stoichiometric YBCO-248, which have roughly one-order-of-magnitude sharper lines; and stoichiometric YBa$_2$Cu$_3$O$_{7}$,
which is sharper by close to two orders of magnitude.

Turning to direct evidence of an inhomogeneous superconducting gap,  superconducting puddles have been  visualized using STM in overdoped Bi2201 and Bi2212 (refs \cite{gomes_visualizing_2007,scalapinoaharonandme,abhay1,abhar2,milanpuddles}). At $p = 0.19$, superconducting gaps in Bi2212 persist up to 120 K 
or more in local regions with radii of a few tens of angstroms. A similar temperature scale characterizing the onset of some form of local superconducting correlations has been inferred from analysis of the Nernst effect and magnetization \cite{wang_nernst_2006}.  ARPES measurements on the same material show a particle--hole symmetric (so presumably superconducting) gap that opens at an imprecisely defined crossover temperature that remains of order 50\% above \Tc for \Tc as low as 70 K (refs \cite{sudichen2019,yuhe2021,He_2021,he_rapid_2018}).

Bulk measurements \cite{wang_weak-coupling_2007} 
have clearly established the existence of a   $\sqrt{H}$ dependence of the specific heat (where $H$ is magnetic field strength) down to low fields in overdoped LSCO.
They also showed that the residual, $T\to 0$, value of the Sommerfeld coefficient, $C/T$, in LSCO at several overdoped values of $p$ is more than half of its normal-state value at $T> \Tc$.  
To obtain such a large residual $C/T$ from conventional theories of pair breaking would require a normal-state scattering rate $\Gamma$ of order $\hbar \Gamma =1$ meV. But this same scattering rate would produce $H$-linear specific heat below approximately 4 T, which is clearly at odds with the observation of $C/T \propto \sqrt{H}$ down to 0.5 T.  To reconcile these observations, it has been suggested \cite{wang_weak-coupling_2007} that LSCO is heterogenous at these dopings, with relatively clean regions that become superconducting and show $C/T \propto \sqrt{H}$, and non-superconducting regions that 
account for the residual specific heat. As outlined above, 
this proposal is broadly consistent with theoretical expectations for $d$-wave superconductivity with a short coherence length in the presence of randomly distributed dopant atoms. It is also consistent with optical and THz spectroscopy on overdoped LSCO and Ca-YBCO, which demonstrate a large fraction of uncondensed carriers in a Drude-like peak below \Tc (refs \cite{tajima_-plane_2005,uykur_-plane_2011,mahmood_effect_2022}).  It is worth noting that a similarly large uncondensed fraction of quasiparticles is also present in the specific heat of overdoped Tl2201: by $\Tc = 30$ K, the jump in $C/T$ at \Tc is only 10$\%$ of the residual $C/T$ as $T\rightarrow 0$ (ref. \cite{wade_electronic_1994}).

Further evidence suggesting that pairbreaking in an effective-medium approximation fails to account for the suppression of \Tc at the edge of the dome comes from irradiation studies on LSCO (ref. \cite{mahmood_effect_2022}) and YBCO (ref. \cite{rullier-albenque_influence_2003}). There is also bulk evidence of strongly electronically inhomogeneous states produced by alloy disorder in LSCO (refs \cite{tranquadagrains} and \cite{tranquadastripepercolation}).

\begin{figure}
	\centering
	\includegraphics[width=\linewidth]{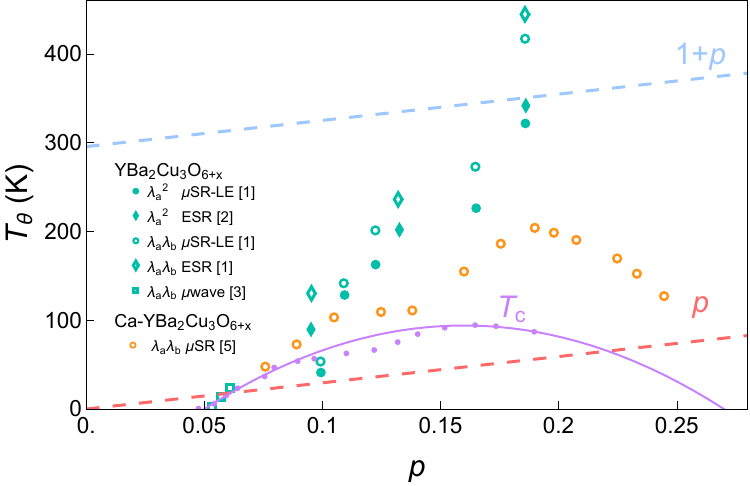}
	\caption{\textbf{Doping dependence of the phase stiffness.}
		Here the measured values of $T_\theta$, the energy scale that characterizes the prevalence of thermal phase fluctuations --- defined in equation \ref{Ttheta} 
		for the case in which $L=a_c$, the mean distance between Cu--O planes --- is shown as a function of the hole doping $p$ for \ybco{6+x} (teal symbols) and Ca-doped \ybco{6+x} (gold symbols).
		Although \Tc (shown for \ybco{6+x}, purple line) at fixed doping is only weakly affected by Ca substitution, the Ca-doped materials exhibit a clear suppression of the superfluid stiffness with increasing hole doping compared with the Ca-free material. Different ways of measuring the penetration depth, $\lambda$, give somewhat different values.  Moreover, because \ybco{} is orthorhombic, the penetration depth  is different along the $a$ and $b$ directions (perpendicular and parallel to the Cu--O `chains', respectively), so different ways of averaging over  $\lambda_a$ and $\lambda_b$  produce  different results. The dashed lines show the values of $T_\theta$ expected in a clean system of electrons with effective mass $m^*=5 m_e$ (a representative value extracted from quantum oscillations in overdoped Tl2201; $m_e$ is the bare electron mass), under the assumption that the density of electrons is $1+p$ (blue dashed line) or $p$ (pink dashed line) per planar Cu. (There are two planar Cu atoms per unit cell in the \ybco{7} crystal structure.) Low-energy muon-spin-resonance data ($\mu$SR-LE) are taken from refs \cite{baglo_detailed_2014,kieflDirectMeasurementLondon2010}; electron-spin-resonance (ESR) data from ref. \cite{pereg-barnea_absolute_2004}; microwave data from ref. \cite{brounSuperfluidDensityHighly2007}; and $\mu$SR data on Ca-YBCO from ref. \cite{bernhardAnomalousPeakSuperconducting2001}.
	}
	\label{fig:phase_stiff_all}
\end{figure}

\ybco{7} stands alone as a material that becomes more homogeneous as it is doped from optimal to overdoped \cite{hosseini_microwave_1999,turner_observation_2003}, owing to its particular mechanism of doping through copper-oxide chains. Where Tl2201, LSCO, Bi2212, and even Ca-doped YBCO generally have phase stiffnesses that decrease 
when moving from optimal doping to overdoping, \ybco{7}'s phase stiffness increases substantially (\autoref{fig:phase_stiff_all}).
Our specific prediction is that \ybco{7} doped further using pressure or uniaxial strain would show a continued rise in superfluid phase stiffness.

Finally, we note that there are reasons to believe that the small values of $T_\theta$ seen in underdoped cuprates reflect distinct, more likely intrinsic physics than in the overdoped regime. (This does not imply that disorder plays no role in the behaviour of the underdoped material --- but it does suggest that an underlying piece of strong-correlation physics is the primary driver of the small values of $T_\theta$.) Specifically, there is independent evidence that not only $T_\theta$, but more generally the density of mobile charges drops with underdoping. For instance, the Drude weight,
$D$, obtained by integrating the normal-state optical conductivity up to frequencies of several hundred meV,  
drops on the underdoped side upon approach to the Mott insulating state, but it remains large upon approaching the overdoped side of the superconducting dome; on the underdoped side, the fraction of mobile carriers that condense in the superconducting state ($\sim T_{\theta}/D$) remains nearly constant, while on the overdoped side, the drop in $T_\theta$ is associated with a decreasing fraction of condensing carriers \cite{armitagemissingweight,michonSpectralWeightHoledoped2021}.

\subsection*{The weak-coupling magnitude of the superconducting gap}

Weak-coupling BCS theory makes a specific prediction for the ratio of the superconducting gap to the critical temperature. For a $d$-wave superconductor, this ratio depends weakly on band-structure details, and a typical value is $\Delta/k_{\rm B}\Tc=2.14$ (ref. \cite{musaelian_mixed-symmetry_1996}). 

\subsubsection*{Concerning the BCS value of $\Delta_0/k_BT_c$.}
ARPES and STM measurements of the superconducting gap in Bi2212 show that $\Delta/k_{\rm B}\Tc$ can be as  
large as 8 in the underdoped regime \cite{vishik_phase_2012,gomes_visualizing_2007}. This is indicative either of strong coupling, or of the difficulty in separating the pseudogap from the superconducting gap. However, even if only the near-nodal gap \cite{vishik_phase_2012} is assumed to be the direct consequence of pairing, this ratio grows with underdoping and reaches values in excess of 4.  We will return to this later when discussing the shape of the $d$-wave gap in momentum space, but for now we note that this is sufficient evidence to conclude that underdoped cuprates clearly do not follow the simple weak-coupling prediction for the size of the gap.

When Bi2212 is hole-doped past optimal doping, however, both STM and ARPES show that the gap decreases rapidly and the ratio $\Delta_0/\Tc$ approaches 2 (within substantial error bars) by $p = 0.24$ (refs \cite{he_rapid_2018,gomes_visualizing_2007,drozdov2018phase}). Furthermore, although STM shows a broad distribution of gap sizes ---  
and correspondingly some local regions in which a gap persists to temperatures well above \Tc --- the mean value of $\Delta_0/2$ and the typical temperature at which the local gap vanishes 
both approach \Tc 
for $p \gtrsim 0.24$.  All of these observations suggest that the strength of the superconducting gap in overdoped cuprates is roughly consistent with weak-coupling BCS theory. 

\subsubsection*{Concerning the non-mean-field $T$-dependence of the gap.}
Although the persistence of superconducting gaps above \Tc --- seen in both STM and ARPES --- is inconsistent with the BCS mean-field assumption, as already discussed it is natural to assign this to the intrinsic amplification of the effects of mesoscopic variations in the local environment expected in a $d$-wave superconductor with relatively short coherence length.

\subsection*{Absence of intertwined orders}
A central feature of the cuprate phase diagram is the presence of multiple intertwined phases that all occur with comparable energy scales --- antiferromagnetism, spin glass, charge density wave
and $d$-wave superconductivity all appear at the scale of a few tens to a couple of hundred kelvin \cite{fradkin_colloquium_2015}. (If the pseudo-gap can be associated with some other form of `hidden order' \cite{chakravarty_hidden_2001}, then it too must be characterized by this same general energy scale.)  The lack of separation of energy scales makes it impossible to cleanly treat any single order parameter in isolation \cite{laughlin_hartree-fock_2014,zhang_unified_1997}.  

The situation is greatly simplified in the overdoped regime. There, the various  
non-superconducting orders clearly weaken and plausibly disappear beyond a doping that depends on the particular order under consideration, as well as on the particular cuprate. For example, charge-density-wave order (with correlation lengths $\xi_{cdw} \lesssim 200$ \AA) is present in Tl2201 up to $p = 0.25$, but is absent at $p = 0.26$ and above. This is apparent from both resonant inlelastic X-ray scattering at $p = 0.25$ and $p = 0.26$ (ref. \cite{tam_charge_2022}), as well as from the absence of Fermi-surface reconstruction in quantum-oscillation experiments at higher dopings \cite{rourke_detailed_2010}. 

Likewise in YBCO, long-range antiferromagnetism is absent above $p = 0.05$, and short-range spin glass correlations are absent above $p = 0.1$. Above that doping, a gap opens 
in the spin excitation spectrum seen in neutron scattering \cite{hinkov_spin_2007}. For fully doped YBCO ($p = 0.19$), the gap in the spin excitation spectrum is of order 40 meV (ref. \cite{fong_polarized_1996}), implying that static spin structures (antiferromagnetism and related orders) and low-energy spin excitations are not relevant at these dopings. No spin order, either short- or long-range, has been reported in Tl2201. 

Although there is no doubt that the various forms of spin-density-wave and charge-density-wave stripe order seen in LSCO are strongest in underdoped LSCO, with increasingly detailed studies it has become clear that weak correlations --- with much the same stripe structure --- persist deep into the overdoped range of $p$, and possibly to the edge of the superconducting dome.  However, careful analysis suggests that this order exists in the form of strongly correlated striped patches in an otherwise metallic matrix \cite{tranquadagrains}.  This observation, in turn, probably reflects a similar sort of correlation-enhanced amplification of the effects of otherwise innocuous mesoscopic variations in the local environment --- the density-wave analogue of the explanation offered for the existence of superconducting puddles.

\section*{The nature of the mechanism}
Accepting the validity of the BCS perspective in the overdoped cuprates, it is possible to work backwards from observations of the gap function to draw incomplete but clear inferences on the nature of the interactions responsible for the pairing. We focus on the ARPES results on Bi2212, where good tight-binding fits to the electronic dispersion have been obtained and for which the cleanest measurements of the superconducting gap are available.

For doping concentrations ranging from slightly underdoped to the highest level of overdoping for which data exist, and within the error bars, the $\vec k$ dependence of the $T\to 0$ gap function on the Fermi surface is given simply by $d(\vec k)=(1/2)\left[\cos k_x - \cos k_y\right]$ --- that is, $f(|\vec k|) \approx 1$ in equation \ref{equation6}. As we have already discussed, this implies that the pairing interaction is short-ranged, acting primarily on nearest-neighbours.  
(Given the importance of this inference, it would be desirable to obtain more quantitative experimental bounds on the magnitudes of any higher gap harmonics. It 
would also be desirable to explore the frequency dependence of the gap function, but this is difficult without a better understanding of the full spectral function.)

\begin{figure}
	\centering
	\includegraphics[width=1.0\linewidth]{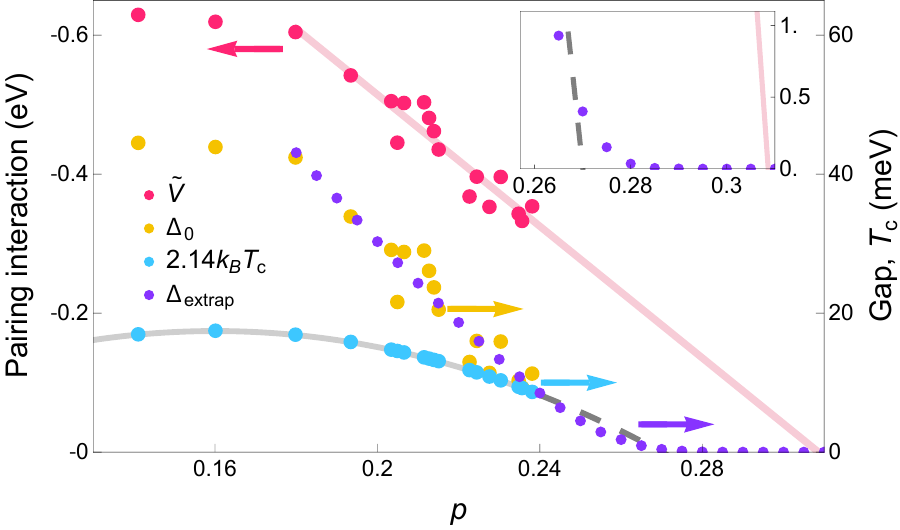}
	\caption{\textbf{Effective pairing interaction in Bi2212.} Inferred from angle-resolved photoemission spectroscopy (ARPES) data \cite{he_rapid_2018}, the measured values of the quasi-particle gap magnitude, $\Delta_0$, (teal points) are compared with the expected BCS magnitude of $\Delta_0=2.14k_\textrm{B}\Tc$ (dark blue points) for the measured transition temperature \Tc. Where BCS theory is applicable, these points should approximately coincide. The values of the pairing interaction $\tilde V$ (red points) are obtained by inverting equation \ref{linearizedgap} using the measured values of $\Delta_0$ and the normal-state dispersion.  Simple fits to the  existing data (black and red lines) enable smooth extrapolations to higher values of doping, for which no data exist. The purple points ($\Delta_\textrm{extrap}$) are the values of $\Delta_0$ that would result from values of $\tilde V$ on the red line.  The inset is an enlargement of the doping range around the critical point --- at which the extrapolated $\tilde V$ changes sign. 
	}
	\label{fig:gapfigure}
\end{figure}

Inverting equation \ref{equation6} without introducing any additional high-energy cutoff (taking $F(\epsilon)=1$), the magnitude of the pairing interaction  can be determined from an empirical fit to the measured dispersion and the gap function.  For the relevant range of dopings, the ARPES dispersion is well fit by 
\begin{align}
	\epsilon(\vec k) =& - 2t [\cos(k_x)+\cos(k_y)] 
	-4t' \cos(k_x)\cos(k_y)\nonumber \\
	&- 2t'' [\cos(2k_x)+\cos(2k_y)], 
	\label{bandstructure}
\end{align}
with $t=166$ meV, $t' = -0.33 t$ and $t''= 0$ \cite{chen_unconventional_2022}, and with a bilayer splitting of the form $-2t_\perp d^2(\vec k)$ with $t_\perp= 0.21 t$ \cite{ai_distinct_2019}. The measured values of the gap (averaged over the bonding and anti-bonding bands) and the value we extract for the pairing strength, $\tilde V$, are shown in \autoref{fig:gapfigure}.  

We can make several inferences from analyzing $\Delta(\vec{k})$. The magnitude of the pairing interaction at optimal doping,  $\tilde V \approx 0.6$ eV, is of the expected magnitude if it is assumed that it derives primarily from local exchange interactions.  Specifically, it is well known that, to next to leading order in the large $U\equiv \tilde U(0)$ expansion of the pure Hubbard model, an effective nearest-neighbour $J \equiv \tilde J(1) = 4t^2/\tilde U$ is generated, and $\tilde U(1) = -\frac 14 J$.  For the inferred value of $\tilde V=3\tilde J(1)-4\tilde U(1) =4J$,  this corresponds to  $J\approx 0.15$ eV, in reasonable agreement\cite{andersontirade}  with values inferred in  undoped antiferromagnet cuprates \cite{valuesofJ}. A phenomenological analysis of the pairing mechanism in optimally doped YBCO has been carried out  using the measured magnetic structure factor as the bosonic `glue' in the Migdal--Eliashberg version of the gap equation \cite{keimer1,keimer2}. This approach gives a mean-field \Tc that is roughly twice the actual \Tc, consistent with the conjecture that both pairing and phase fluctuations are of comparable importance in this crossover regime. Notably, a relatively wide range of energies (of order the width of the paramagnon dispersion, $\Delta E \sim 2J \approx 0.3$ eV) contribute to the pairing interaction. This implies that the frequency dependence of the interactions can be neglected at lower energies (including energies comparable to the superconducting gap) --- that is, the results should be equivalent to those obtained using a BCS approach.

To the extent that $J$ is an extension of the local physics of the undoped antiferromagnet, it would be expected that the effective $J(p)$  decreases with large enough doping. The smooth but none-the-less pronounced decrease of $\tilde V$ with increasing $p$ thus provides additional suggestive evidence that the pairing interaction is largely magnetic in origin.
Comparing this expectation with direct probes of the doping dependence of the magnetic fluctuations is somewhat complicated. X-ray measurements\cite{deanoverdoped,keimeroverdoped,wakimotoX} suggest that  there is relatively little  decrease in the strength of the magnetic fluctuations  over much of the  Brillouin zone (especially near the scattering wavevector $\vec Q=(\pi,0)$). However, neutron scattering\cite{wakimoto1,wakimoto2,tranquadagrains} studies have  found that in the important\cite{scalapinooverdoped} range of $\vec Q$ near $(\pi,\pi)$,  there is indeed a pronounced drop in magnetic intensity with overdoping.

The fact that the measured $\vec k$-dependence of the gap is well approximated by $d(\vec k)$ implies that the effective interaction remains short-ranged even 
for $p$ as small as $p_{\textrm{opt}}$. This precludes any theory that identifies optimal doping with the point at which the pairing arises primarily from the exchange of any familiar quantum-critical collective mode. Note that this does not imply that there are no quantum critical points under the superconducting dome --- given the plethora of intertwined orders, such quantum critical points may well arise in some circumstances. (For instance, the observation\cite{ramshawquantum oscillation} of an apparently divergent electron effective mass seen in quantum oscillations in YBCO as $p$ approaches a critical value $ \sim p_{opt}$ is strong evidence of the existence of a quantum critical point in the `normal' state, obtained when superconductivity  is quenched with a high magnetic field.) However, to the extent that $\Delta(\vec k) \propto d(\vec k)$, we  conclude that no quantum critical mode is the prime driver of superconductivity. 

From this viewpoint, it is interesting to compare the $\vec k$-dependence of the gap in the hole-doped cuprates with the corresponding (not nearly as well documented) behaviour of the electron-doped cuprates.   In the electron-doped materials,  even at optimal doping there is direct evidence of antiferromagnetic fluctuations with a correlation length that is large compared to the lattice constant. Given that these fluctuations play a role in the pairing interaction, a much more structured $\Delta(\vec k)$ would be expected: indeed, suggestive evidence of a non-monotonic dependence of the gap along the Fermi surface has been observed (see ref. \cite{greenereview} for a review). 

It is important to recognize that, at short distances and high energy scales, it follows  from the low electron density (large value of $r_s$) that the cuprates are  more strongly interacting than conventional metals.  The above treatment of the mechanism of superconductivity in  overdoped cuprates does not contradict this  observation.  The essential assumption we are making is that when high-energy degrees of freedom are integrated out, the quasiparticles in the resulting low-energy effective Hamiltonian are sufficiently weakly interacting for a Fermi-liquid description to be possible; both the parameters that characterize the quasiparticle dispersion relation in equation \ref{bandstructure} and the effective pairing interaction, $\tilde V$, should thus be interpreted as   effective parameters. 

The self-consistency of this analysis can legitimately be questioned, given that we have obtained values of the pairing interaction, $\tilde V$, that are not  small: 
$\tilde  V \rho(E_F) \sim 1$ at optimal doping, dropping to  
$\tilde  V \rho(E_F)\sim 0.5$ at the highest doping (\autoref{fig:gapfigure}).
However, the same can be said of the Migdal--Eliashberg treatment of conventional superconductors such as  Pb, for which the dimensionless strength of the electron--phonon coupling  $\lambda \sim 1$ (ref. \cite{allenreview}).

\section*{Experimental tests}

Although it is not  possible for a material to be entirely free from disorder, it is possible to identify circumstances in which the amount of disorder is greatly suppressed. The principal empirical test of the viewpoint we have presented here is that deviations from the predictions of Fermi-liquid theory for the low-energy properties of the normal state, and of BCS mean-field theory for the superconducting state, become less pronounced in overdoped cuprates when disorder is weaker. 

There are various experimental properties that could serve as particularly good indicators of such trends (and already some encouraging evidence of the expected tendencies):

\begin{itemize} 
	\item With decreasing disorder, the linear drop of $\Tc\sim (p_{\textrm{max}}-p)$ to zero at a critical doping $p_{\textrm{max}}$ should become increasingly rounded (convex) and $p_{\textrm{max}}$ is likely to increase (\autoref{fig:gapfigure}). 
	The fact that \Tc persists up to at least $p = 0.31$ in Tl2201 \cite{rourke_detailed_2010}, compared to $p = 0.27$ in the comparatively more disordered LSCO, is suggestive of this, but a finer doping dependence is needed.
	
	\item For fixed $p$, the superfluid density should grow with decreasing disorder to be an increasingly large fraction of the total Drude weight. Specifically, there should be an increasing tendency towards the edge of the superconducting dome to realize the familiar situation in which $T_\theta \gg \Tc$. 
	Data for slightly-overdoped \ybco{x} suggest that this may be the case in the cleanest cuprates, but higher dopings are needed.
	
	\item In the cleanest materials, the zero-temperature gap function should approach the BCS value in magnitude, $\Delta_0 \approx 2\Tc$. Moreover, spectroscopic evidence for a gap that persists for $T > \Tc$, or beyond the end of the superconducting dome, $p > p_{\textrm{max}}$, should occur over decreasing ranges of $T$ and $p$. This is suggested in existing data for Bi2212 (\autoref{fig:gapfigure}), but higher dopings are needed.
	
	\item The specific heat should show increasingly clear mean-field features with decreasing disorder, including a mean-field jump in the specific heat at $\Tc$ and a strong suppression of the residual $C/T$ in the  $T\to 0$ limit.  There should be a similar drop in the residual Drude response seen in terahertz optical conductivity for $\omega \ll \Delta_0$ and $T\ll \Tc$. 
	
	\item  The $T$-linear term in the resistivity, seen in overdoped materials when superconductivity is suppressed by a magnetic field, should become increasingly small with decreasing disorder.
	
\end{itemize}

There are promising indications that the properties of nearly `ideal' underdoped cuprates can be explored by focusing on  the inner planes of multilayer cuprates.  Concerning the principle focus of this article, \ybco{7} is somewhat overdoped and is stoichiometric, meaning that intrinsic substitutional disorder is absent. Indeed, various experimental probes seem to corroborate the expectation that available samples of \ybco{7} are tangibly less electronically inhomogeneous than other overdoped materials. 
For the future, it seems that the doped hole concentration in the Cu--O planes of this material can be increased with application of pressure --- so this might be the ideal material with which to test our ideas.

\section*{Acknowledgments}

We would like to acknowledge extremely useful discussions with   P. Armitage, B. Atkinson, M. Beasley,  I. Bozovic, S. Brown, S. Chakravarty, A. Chubukov, R. Greene, H. Guo, S. Hartnoll, P. Johnson, M.-H. Julien, B. Keimer, R. B. Laughlin, A. Millis,  M. Norman, S. das Sarma, J. Schmalian, M. Sulangi, S. Uchida,    T. Valla,  and H.-H. Wen  and particularly detailed and insightful feedback from Y. He, J. Tranquada, P. Hirschfeld, and D. Broun.  S.A.K. was supported in part by NSF-BSF award DMR-2310312. B.J.R. was supported in part by NSF award DMR-2428169 and the Canadian Institute for Advanced Research.

\section*{Author Contributions}

B.J.R and S.A.K contributed equally to all aspects of the project. 

\section*{Competing interests} The authors declare no competing interests. 

\section*{Data and Code availability}
All data are available from the authors upon request.


\end{document}